\documentclass{ws-p9-75x6-50}

\begin{document}

\title{Renormalization Group Flow Equations for the scalar $O(N)$ theory\footnote{Talk given at the Second Conference On The Exact
Renormalization Group, Rome, Italy, September 18-22, 2000}}

\author{B.-J. Schaefer,\footnote{schaefer@nhc.tu-darmstadt.de}\\[1ex]}


\author{O. Bohr and J. Wambach}

\address{Institut f\"ur Kernphysik, Schlossgartenstr. 9, D-64289 Darmstadt, Germany}

\maketitle

\abstracts{Self-consistent new renormalization group flow equations for
an $O(N)$-symmetric scalar theory are approximated in next-to-leading
order of the derivative expansion. The Wilson-Fisher fixed point in
three dimensions is analyzed in detail and various critical exponents
are calculated.}

\section{Introduction and conclusion}

In this talk we present novel nonperturbative flow equations for an
$O(N)$-symmetric scalar theory in next-to-leading order of the
derivative expansion. Based on the Wilsonian Renormalization Group
method also known as the Exact Renormalization Group (ERG) approach we
combine a Schwinger proper time regularization scheme with a
perturbative treatment in order to obtain self-consistent RG flow
equations. We investigate in detail the nontrivial Wilson-Fisher fixed
point in three dimensions and determine universal quantities such as
the critical exponents $\eta$, $\nu$, $\beta$ and $\delta$
independently for arbitrary numbers of field components $N$ without
resorting to any polynomial truncations. For this approach the typical
and inherent analytical threshold functions in the flow equations,
which describe the continuous decoupling of the massive modes from the
RG evolution enable a direct and smooth linkage between the
four-dimensional system at zero temperature and the critical $O(N)$
universal behavior near the critical temperature $T_c$. The
dimensional reduction phenomenon is embedded in the threshold
functions in a very transparent way.\cite{wetter,bohr} The prediction
of the universal critical exponents is one of the great achievements
of the ERG methods but they may also be applied to other systems
e.g.~with a first-order phase transition.\cite{tetra} All ERG
equations are equivalent up to the choice of the regulator or cutoff
function. For instance Polchinski's equations in the limit of a sharp
momentum cutoff are the Wegner-Houghton equations and are equivalent
to Wilson's equations.\cite{wegner} In order to extract physical
relevant quantities from the ERG equations which are complicated
highly non-linear functional equations one has to employ an
appropriate approximation or truncation scheme. The ERG equations
themselves are by construction scheme independent but this need not be
true any longer in specific truncations. Thus, different realizations
of the cutoff function or regulator for the ERG equations yield an
uncertainty in the determination of universal quantities. To leading
order the critical exponents are scheme inpedendent but in
next-to-leading order the results become scheme dependent. This makes
it particularly difficult to determine uniquely the critical exponent
$\eta$. For instance in four dimensions the sharp cutoff
Wegner-Houghton equation predicts the proper field anomalous dimension
$\eta$ in next-to-leading order in the derivative expansion but it
fails at the Wilson-Fisher fixed point in three
dimensions.\cite{zappa} We propose a smooth cutoff version of the ERG
based on a heat-kernel regularization which should overcome these
difficulties and obtain remarkable precise values for the anomalous
dimension $\eta$ in a very simple uniform wavefunction renormalization
approximation by neglecting the field dependence in $Z_k$. In order to
provide an estimate of the scheme dependence we determine several
critical exponents for different implementations of the smooth cutoff
functions and find a rapid convergence to stable values that are in
good agreement with entirely different approaches. Another criterion
for estimating the scheme dependence may be based on the minimum
sensitivity principle which we do not further pursue in this
work.\cite{litim} The next section is devoted to a brief review of the
derivation of the flow equations for the potential $U_k$ and the
coefficient $Z_k$ of the lowest-order derivative term.\cite{vincenzo}

\section{The flow equations with a proper time regularization}

As an illustrative example and as a test we apply the method with a
proper time regulator to an $O(N)$ scalar theory. In the ultraviolet
(UV) region of the theory at the scale $\Lambda$ of the order of $1$
GeV we consider the following effective Lagrangian
\be
{\cal L}_\Lambda = \frac{1}{2} \left( \partial_\mu \vec{\Phi} \right)^2 + 
\frac{\lambda}{4} \left( \vec{\Phi}^2 - \Phi_0^2 \right)^2
\ee
with the $N$-component vector $\vec{\Phi} = (\Phi_1, \Phi_2, \ldots ,
\Phi_N )$. The field $\Phi_0^2$ denotes the minimum of the
potential. The expansion in powers of momenta of the effective action
up to order ${\cal O} (\partial^4)$ reads
\be
\label{actionon}
\Gamma [\vec{\Phi}] =\int d^d x \left\{ -U(\vec{\Phi}^2 ) +\frac 12
Z_1 \left( \vec{\Phi}^2 \right)
\left( \partial_{\mu} \vec{\Phi} \right)^2 + \frac{1}{2} Z_2 \left(
\vec{\Phi}^2  \right)
\left(\vec{\Phi} \partial_{\mu} \vec{\Phi} \right)^2
\right\} \ .
\ee
The perturbative one-loop contribution to the effective action yields
formally a non-local logarithm which we regularize by a proper-time
regularization resulting in a finite local action. We implement a
multiplicative, a priori unknown, blocking or smearing function
$f_k(\tau)$ which governs the coarse-graining in the proper-time
integrand. In this way the sharp proper-time cutoff is replaced by a
smooth one and the scale $k$ acts as an infrared cutoff separating the
low- and high-momentum modes of the fields. After introducing a
complete set of plane wave states in the heat kernel this procedure
yields the following $d$-dimensional effective action\cite{heat}
\be
\label{gammaheat}
\Gamma [\vec{\Phi}] = -\frac 12 \int d^d x \int_0^{\infty} \frac{d \tau}{\tau} f_k
\int 
\frac{d^d p}{(2\pi)^d} \mbox{tr} \ e^{- \tau 
\left( p^2 - 2ip_{\mu} \partial_{\mu} - \partial^2
+V_{ij}'' (\Phi)\right)}
\ee
with the shorthand notation $V_{ij}'' (\Phi ) = \frac{\delta^2
V}{\delta\Phi_i \delta \Phi_j}$ for the $(N \times N)$-matrix valued
second derivative of the $O(N)$-symmetric potential $V$ at the UV
scale $\Lambda$. The second derivative is given by $V_{ij}''=
\lambda \left( \vec{\Phi}^2 -\Phi_0^2 \right) \delta_{ij} + 2
\lambda \Phi_i \Phi_j$. The trace in Eq.~(\ref{gammaheat}) runs over
the $N$ fields and can be evaluated analytically with standard
techniques. In order to find the next-to-leading order of the
effective action we expand $\exp \left[ -\tau \left( -2ip_\mu
\partial_\mu -\partial^2 + V'' \right) \right]$ in powers of
derivatives up to second order. To this order the action separates
into two parts
\be
\Gamma_2 = \Gamma^{(0)} + \Gamma^{(2)}\nonumber
\ee
with $\Gamma^{(0)}$ denoting the one-loop potential contribution
(containing no derivatives) and $\Gamma^{(2)}$ the second-order
contribution (containing two derivatives). Comparing the expansion
coefficients $U$, $Z_1$, $Z_2$ of the effective action
(\ref{actionon}) with the corresponding terms in $\Gamma_2$ we can
extract the effective potential contribution and the wavefunction
renormalization contributions. In order to simplify the work
drastically we perform a uniform wavefunction renormalization by
neglecting the field dependence and consider only one wavefunction
renormalization $Z_1=Z_k$. It turns out that this simple approximation
already describes all qualitative features at the phase
transition. The desired self-consistent flow equations are obtained by
differentiation of the equations with respect to the infrared scale
$k$ followed by the substitution $V \to U_k$ on the right-hand
side. This replacement has an analogy to the Schwinger-Dyson
self-consistent resummation.  The structure of the flow equations
depends on the choice of the blocking functions but the universal
physical results which are obtained by solving these flow equations
towards the infrared should not depend on the specific choice. The
regulator essentially governs the way how the irrelevant operators of
the theory are integrated over.  Here we chose for any integer $M\geq0$
the following form for the derivative of the blocking functions
\be\label{smearfkt}
k \frac{\partial f_k^{(M)} (\tau)}{\partial k} \sim -2 (\tau Z_k
k^2)^{(M+2)} e^{-\tau Z_k k^2}\ .
\ee
and obtain e.g. for the blocking function with $M=1$ the rescaled
dimensionless flow equations in $d$ dimensions
\begin{eqnarray}
\label{flowpotcrit}
k\frac{\partial u_k({\phi}^2)}{\partial k} &=& - d u_k +(d-2+\eta ){\phi}^2
u'_k +
\frac{S_d}{d} \left[ \frac{1}{1+2u'_k +4{\phi}^2 u''_k}
+\frac{N-1}{1+2u'_k}
\right]
\end{eqnarray}
\begin{eqnarray}
\label{flowzcrit}
- \frac{k}{Z_k} \frac{\partial Z_k}{\partial k} &=&  \left. \frac{2 S_d}{{\phi}^2 d} 
\left[ 1 + 
\frac{1}{ (1+4 {\phi}^2 u''_k)^2 } + \frac{1}{2 {\phi}^2 u''_k}
\left( 
\frac{1}{1+4 {\phi}^2 u''_k} - 1 \right) \right] \right|_{\phi=\phi_0} \ ,
\end{eqnarray}
\begin{eqnarray}\nonumber
\mbox{where}\qquad  S_d = 2/\Gamma(d/2)(4\pi)^{d/2} ,\  {\phi}^2 &:=&
k^{-(d-2+\eta)} \Phi^2_k\quad   \mbox{and}\quad  u_k({\phi}^2) := k^{-d }
U_k({\phi}^2)\hfill\ \ 
\end{eqnarray}
and we have introduced the anomalous dimension $\eta = -k\partial_k
\ln Z_k$. A prime on the potential denotes differentiation with
respect to ${\phi}^2$.  Because we have omitted the field dependence
of the wavefunction renormalization the flow equation
(\ref{flowzcrit}) for $Z_k$ has to be evaluated at the minimum of the
potential $\phi_0$.

\section{Results}

\begin{figure}[t]
\epsfxsize=16pc 
\centering 
\epsfbox{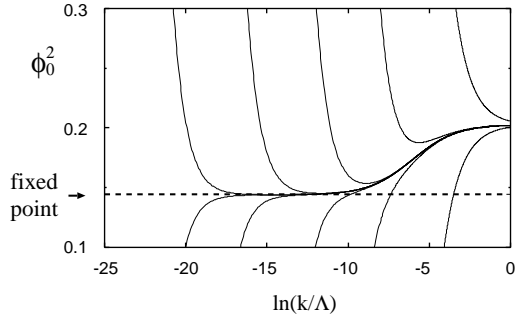} 
\caption{The evolution of the dimensionless minimum towards the
scaling solution.  \label{fig1}}
\end{figure}

In order to test the approximations described above we solve the
coupled system of flow equations by discretization of the field
$\phi^2$ for a general, not truncated potential $u_k(\phi^2)$
numerically on a grid by a fifth-order Runge Kutta algorithm for each
grid point. Details can be found in the work by O.~Bohr et
al.\cite{bohr}. The flow equations (\ref{flowpotcrit}) and
(\ref{flowzcrit}) are in a scale-independent form and, due to the
dimensional reduction phenomenon, we do not need to use the finite
temperature version of the flow equations at the critical
temperature. We can directly link the $O(N)$ universal behavior near
$T_c$ with the physics at zero temperature and therefore it is
sufficient to employ only the three-dimensional zero-temperature
equations in order to investigate the critical regime of the phase
transition. In this way we circumvent the evaluation of Matsubara sums
for the investigation of the critical point. A second-order phase
transition involves an infrared fixed point of the RG
transformations. Thus, the physics close to the phase transition is
scale invariant and the critical behavior should be described by a
scale-independent solution. This is demonstrated in Fig.~\ref{fig1}
where the $k$-evolution towards zero of the dimensionless minimum
$\phi^2_0$ of the potential for different initial values near the
critical value at the UV scale $k=\Lambda$ is shown. The dashed line
indicates the $k$-independent constant (fixed point) scaling solution
while the evolution near the critical value deviates either towards
the spontaneously broken ($\phi^2_0 \neq 0$) or the symmetric
($\phi^2_0 = 0$) phase. Of course, not only the minimum
$\phi^2_0$ but all quantities show a scaling behaviour.
\begin{figure}[htb]
\epsfxsize=16pc 
\epsfysize=11pc
\centering 
\epsfbox{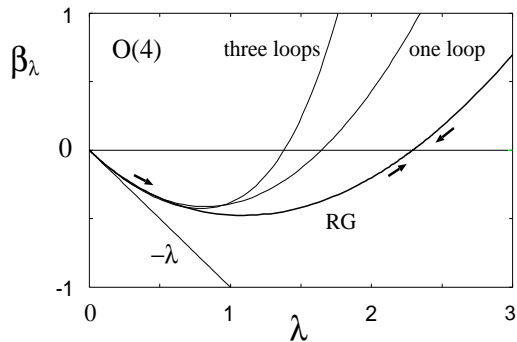} 
\caption{$\beta_\lambda (\lambda)$ for the $O(4)$-model compared to a 
one- and three-loop $\epsilon$-expansion. The arrows indicate the flow
w.r.t. the scale $k$ towards the infrared.\label{fig2}}
\end{figure}
For dimensions less than four two fixed points are present in the
system and can be identified by zeros in the beta functions. In
general it is difficult to calculate the zeros since this requires
knowledge of the physics beyond perturbation theory. In
Fig.~\ref{fig2} the curve, denoted by RG, shows the beta function
$\beta_\lambda$ of the quartic coupling for an $O(4)$-model within our
approach. This result is compared with a one- and three-loop
$\epsilon$-expansion. Thus, the nontrivial infrared-stable
Wilson-Fisher fixed point can clearly by identified around $\lambda^*
\sim 2.3$ for $N=4$. 
\vspace{-5ex}
\begin{table}[htb]
\caption{Critical exponents for various $N$. Our results (RG) are
compared with lattice simulations, $\epsilon$-expansion ($\epsilon$)
and a perturbative calculation ($SPS$).\label{tab1}}
\begin{center}
\footnotesize
\begin{tabular}{|c|c|c|c|c|c|c|}
\hline
$\begin{array}{c}\ \\ \ \end{array}$  & $\eta$  & $\delta$ & $\nu$ & $\beta$  \\
\hline\hline
N=0 $\begin{array}{c} \mbox{RG}  \\ \mbox{lattice}  \\ \epsilon \\ SPS
\end{array}$  
& $\begin{array}{c} 0.039 \\ 0.041(25) \\ 0.032(3) \\ 0.027(4) \end{array}$    
& $\begin{array}{c} 4.77  \\ 4.77(14) \\ 4.81(2) \\ 4.84(3) \end{array}$  
& $\begin{array}{c} 0.59  \\ 0.592(3) \\ 0.588(2) \\ 0.588(2) \end{array}$  
& $\begin{array}{c} 0.30  \\ 0.308(6) \\ 0.304(2) \\ 0.302(2) \end{array}$\\  
\hline
N=1 $\begin{array}{c} \mbox{RG} \\ \mbox{lattice} \\ \epsilon \\ SPS 
\end{array}$  
& $\begin{array}{c} 0.044 \\  0.044(31) \\ 0.038(3) \\ 0.032(3) \end{array}$
& $\begin{array}{c} 4.75 \\  4.75(17) \\ 4.78(2) \\ 4.81(2)  \end{array}$
& $\begin{array}{c} 0.64 \\  0.631(2) \\ 0.631(2) \\ 0.630(2) \end{array}$
& $\begin{array}{c} 0.32 \\  0.329(9) \\ 0.327(2) \\ 0.325(2) \end{array}$\\
\hline
N=4 $\begin{array}{c} \mbox{RG}  \\ \mbox{lattice} \\ \epsilon
\end{array}$  
& $\begin{array}{c} 0.037 \\  0.0254(38) \\ 0.03(1) \end{array}$
& $\begin{array}{c} 4.79 \\  4.851(22) \\ 4.82(6) \end{array}$  
& $\begin{array}{c} 0.78 \\  0.7479(90) \\ 0.73(2) \end{array}$
& $\begin{array}{c} 0.40 \\  0.3836(46) \\ 0.38(1) \end{array}$\\
\hline
N=100 $\begin{array}{c} \mbox{RG}  \end{array}$
& $\begin{array}{c} 0.0025  \end{array}$
& $\begin{array}{c} 4.99  \end{array}$
& $\begin{array}{c} 0.99  \end{array}$  
& $\begin{array}{c} 0.49  \end{array}$ \\
\hline\hline
$\begin{array}{c}\ \\ \  \end{array}$\mbox{large-$N$} & 0 & 5      & 1.0    & 0.5     \\
\hline
\end{tabular}
\end{center}
\end{table}
Note, that this value depends on the particular form of the flow
equations encoded in the choice of the blocking function. However, an
expansion e.g.~of the beta function $\beta_\lambda$ in powers of the
coupling yields in three dimensions $\beta_\lambda = -\lambda +
\frac{2(N+8)}{3\pi^2}\lambda^2+{\cal O}(\lambda^3)$ which agrees
with the first two terms of an $\epsilon$-expansion as it
should. Corresponding to the large-$N$ limit the fixed point
$\lambda^*$ tends to zero merging with the trivial Gaussian fixed
point. Linearizing the flow about the Wilson-Fisher fixed point allows
the determination of the critical exponents by means of finding the
corresponding eigenvalues. We have calculated the exponents $\beta$,
$\delta$, $\nu$ and $\eta$ independently, which allows to test the
well-known scaling relations among the critical exponents. Their
values are listed in Tab.~\ref{tab1} for various $N$. The agreement
with different approaches is a further confirmation of our
approximations described above. One realizes explicitly the
convergence of all calculated critical exponents to the large-$N$
values quoted in the last line of Tab.~\ref{tab1}. In the large-$N$
limit the local potential approximation becomes exact and the
anomalous dimension $\eta$ vanishes.

In order to investigate the scheme dependence of the non-truncated set
of flow equations for the full potential we derive flow equations for
the blocking functions defined in Eq.~(\ref{smearfkt}) with $M=0, 1,2$
and neglect the wavefunction renormalization. As an example we take
$N=4$ and calculate again the critical exponent $\nu$ and $\beta$. The
dependence of these critical exponents on the blocking function of
order $M$ is shown in Tab.~\ref{tab2}.
\begin{table}[htb]
\caption{ The critical exponents $\nu$ and $\beta$ for $N=4$ 
with different blocking functions $f^{(M)}_k (\tau )$ (without
wavefunction renormalization).
\label{tab2}}
\begin{center}
\begin{tabular}{|l|c|c|}
\hline
$M$ & $\nu$ & $\beta$ \\
\hline
  0   & 0.853  & 0.42  \\
  1   & 0.815  & 0.405 \\
  2   & 0.814  & 0.403 \\
\hline
\end{tabular}
\end{center}
\end{table}
A small systematic decrease in the values with the order $M$ is
observed. This behaviour is also seen in the results of
Refs.~\cite{zappa,papp} where the dependence of the parameter $M$
practically vanishes for bigger values of $M$. Already for $M\geq1$ we
do not see any difference in the second significant digit. Going
beyond the Local Potential Approximation (LPA) and taking the
wavefunction renormalization into account this tendency becomes more
stable with the order $M$. An analysis of the blocking functions
$f_k^{(M)}$ with increasing $M$ shows that the regulator has the
effect of selecting smaller and smaller momentum shells.\cite{zappa}
It represents for very large $M$-values a kind of sharp cutoff limit
which, however, cannot be directly linked to the sharp cutoff
Wegner-Houghton equation. This means that a deeper understanding of
the relation between this novel approach concerning the heat kernel
cutoff and other formulations of the ERG is still an important
outstanding issue. So far we and other authors have to restrict the
comparisons of the different methods to numerical results
only.\cite{zappa,liao}

\section*{Acknowledgments}

BJS is grateful to V.~Branchina, S.-B. Liao, D.~Litim, J.~Pawlowski,
Ch.~Wetterich and D.~Zappal$\grave{\mbox{a}}$ for valuable discussions
during the workshop. This work has been supported in part by the GSI
and DFG.

\end{document}